\documentclass[12pt]{iopart}

\usepackage{iopams}  

\usepackage{amstext}
\usepackage{graphicx}
\usepackage{color}

\def\Tr{\mbox{Tr}}

\begin{document}


\title[]{Nonequilibrium fluctuations in quantum heat engines: Theory, example, and possible solid state experiments}
\author{Michele Campisi}
\address{NEST, Scuola Normale Superiore \& Istituto Nanoscienze-CNR, I-56126 Pisa, Italy}
\ead{michele.campisi@sns.it}
\author{Jukka Pekola}
\address{Low Temperature Laboratory, O.V. Lounasmaa Laboratory, Aalto University, FI-00076 Aalto, Finland}
\ead{pekola@ltl.tkk.fi}
\author{Rosario Fazio}
\address{NEST, Scuola Normale Superiore \& Istituto Nanoscienze-CNR, I-56126 Pisa, Italy}
\ead{rosario.fazio@sns.it}

\begin{abstract}
We study the stochastic energetic exchanges in quantum heat engines. Due to microreversibility, these obey a fluctuation relation, called the heat engine fluctuation relation, which implies the Carnot bound: no machine can have an efficiency larger than Carnot's efficiency. The stochastic thermodynamics of a quantum heat engine (including the joint statistics of heat and work and the statistics of efficiency) is illustrated by means of an optimal two-qubit heat engine, where each qubit is coupled to a thermal bath and a two-qubit gate determines energy exchanges between the two qubits. We discuss possible solid state implementations with Cooper pair boxes and flux qubits, quantum gate operations, and fast calorimetric on-chip measurements of single stochastic events.
\end{abstract}

\maketitle

\section{Introduction}
The field of non equilibrium quantum thermodynamics has received a large impulse in the last two decades due to the discovery of a number of exact relations which characterise the response of physical (possibly small) systems, to external perturbations, namely applied mechanical forces or thermodynamic forces (e.g.  temperature gradients, and chemical potential gradients). 
\cite{Esposito09RMP81,Campisi11RMP83}

Unlike traditional thermodynamics \cite{Fermi56Book}, which focusses on macroscopic quantities, fluctuation relations focus on their microscopic, fluctuating, counterparts. To exemplify this, consider the two fundamental objects of thermodynamic investigation, work and heat. A macroscopic thermal engine delivers a certain amount of work while withdrawing a corresponding amount of heat from a hot thermal reservoir. There can be variations in these amounts between different cycles, but typically these fluctuations are negligible. However as the machine size scales down, likewise will the work output and heat absorbed scale down. Accordingly, their fluctuations will become more and more relevant. It then becomes useful to investigate the stochastic properties of such fluctuating quantities. Fluctuation relations pose stringent constraints on the statistics of such fluctuating quantities like heat and work, due to the symmetries (in particular time-reversal symmetry) characterising the microscopic motions of atoms and molecules form which they originate.

Fluctuation relations have been reported for both classical and quantum systems \cite{Esposito09RMP81,Campisi11RMP83,Seifert12RPP75,Jarzynski11ARCMP2,Marconi08PREP111}. In fact identical fluctuation relations hold  regardless of whether the same system is regarded as classical or quantum. Despite their formal identity, classical and quantum fluctuation relations are deeply different in the way they can be accessed experimentally. Concerning work, for example, while typically one can measure the fluctuating work applied to a classical nano system, e.g., a stretched RNA molecule, by continuously monitoring a displacement $x$ and its conjugate force $f$ (e.g. extension and tension in the molecule), and obtaining the work as $W= -\int f dx$, \cite{Liphardt02SCIENCE296,Collin05NAT437,Douarche05EPL70} this is typically impossible in a quantum system. In the quantum scenario, the situation is much complicated by the invasiveness of the measurement apparatus which can lead to a collapse of the wave function. The prescription accordingly is to measure the energy of the system twice (at the beginning and end of the forcing protocol), by means of two projective measurements and obtain the work as their difference \cite{Tasaki00arXiv,Kurchan00arXiv,Talkner07JPA40,Talkner07PRE75}.

This two-measurement scheme has proved however very challenging from the experimental point of view  \cite{Huber08PRL101,Briro14arXiv}, so much that it has been realised only very recently \cite{An14arXiv14094485}.  This occurrence has triggered  the proposal of a number of alternative methods.
 One such method proposes to replace the two invasive projections with many less invasive measurements (POVM) carried on a smaller portion of the system \cite{Campisi10PRL105}. This method is particularly well suited for studies of transport induced by gradients of temperature and chemical potential \cite{Campisi10PRL105}. Some experiments already exist which can be explained in terms of these multiple measurements \cite{Utsumi10PRB81,Kung12PRX2}. They regard the full counting statistics of electrons transported through double quantum dot due to an applied chemical potential difference. We shall remark however that in those experiments all quantum coherences are suppressed.

Another very ingenious method, which is particularly well suited for obtaining the work statistics of a driven system, requires a special coupling of the driven system to an ancilla, e.g. a qubit, and replaces the two energy measurements with state tomography of the qubit at the sole final time \cite{Dorner13PRL110,Mazzola13PRL110}. This method is a form of Ramsey interferometry and gives experimental access to the characteristic function of work, namely the Fourier transform of the probability density  function of work. This has led to the first experimental measurement of quantum work statistics ever performed. It has been performed in a liquid-NMR set-up, and has reconstructed the work pdf of a driven two level system \cite{Batalhao14PRL113}. A proposal for implementing the method with  solid state quantum devices has been put forward in \cite{Campisi13NJP15}. The most promising aspect of this method is that it can be used not only to asses the work statistics of closed systems as in the performed experiment, but also of systems which stay in contact with a thermal bath \cite{Campisi13NJP15}. 

Roncaglia et al. \cite{Roncaglia14arXiv} have proposed to couple the system to a quantum pointer, e.g. a spin chain. The coupling is engineered so that that a single final projective measurement of the state of the pointer will contain information about the work performed on the system. Like the interferometric method, this method is best suited  for the measurement of work. Its experimental realisation however appears extremely challenging.

In this work we focus on yet another method which has been discussed recently in \cite{Pekola14NJP15,Hekking13PRL111} and is based on the calorimetric measurement of photon released and absorbed by thermal reservoirs. This quantum calorimeter is currently under development. The method is well  suited for simultaneously measuring both heat and work in a driven quantum system which stays in contact with one or more baths. For this reason it is very promising for the experimental study of the stochastic energy exchange of quantum thermal machines.

Since the seminal work of \cite{Scovil59PRL2}, showing how the three level maser could be understood as a thermal machine,
quantum thermal machines have been widely studied in the literature \cite{Alicki79JPA12,Kosloff84JCP80,Geva92JCP96} and are still under vigorous investigation \cite{Linden10PRL105,Abah12PRL109,Gelbwaser-Klimovsky13PRE87,Jiang13PRB87,Balachandran13PRB87,Brandner13PRL110,Uzdin14NJP16,Zhang14PRL112,Correa14SCIREP4,Mazza14NJP16}, see also the recent review \cite{Benenti13arXiv13114430} and references therein. However, while so far the focus was on the average value of heat and work, here we focus on their fluctuations as well. As recently reported \cite{Campisi14JPA47}, a special form of the fluctuation relation holds for quantum thermal machines. This form implies that no quantum thermal engine can over-perform the Carnot efficiency. This universal and exact result was anticipated long ago in \cite{Alicki79JPA12} but only for those quantum mechanical open systems whose dynamics can be well approximated by a Markovian master equation in Lindblad form.

After revisiting the heat engine fluctuation relation, we shall introduce a model of thermal engine based on two-qubits each coupled to its own reservoir and subject to a unitary gate operation. We will identify the regimes when the engine works as heat engine, refrigerator, or heater (dud engine), and study its full stochastic characteristics, including the probability density function of its efficiency. 
The most intriguing features of the presented machine are (a) that at maximum power it can reach efficiency above the Curzon-Albhorn efficiency, and (b) that increasing the speed of its operation increases the power output without affecting its efficiency.

It is important to stress that the engine presented here can be implemented in a real solid state device and its stochastic energetic exchanges can be measured using the current and soon available technology. Below we discuss possible solid state implementations based on the calorimetric measurement scheme.

\section{The Heat Engine Fluctuation Relation (HEFR)}
\label{sec:HEFR}
\begin{figure}
		\begin{center}
		\includegraphics[width=0.5\linewidth]{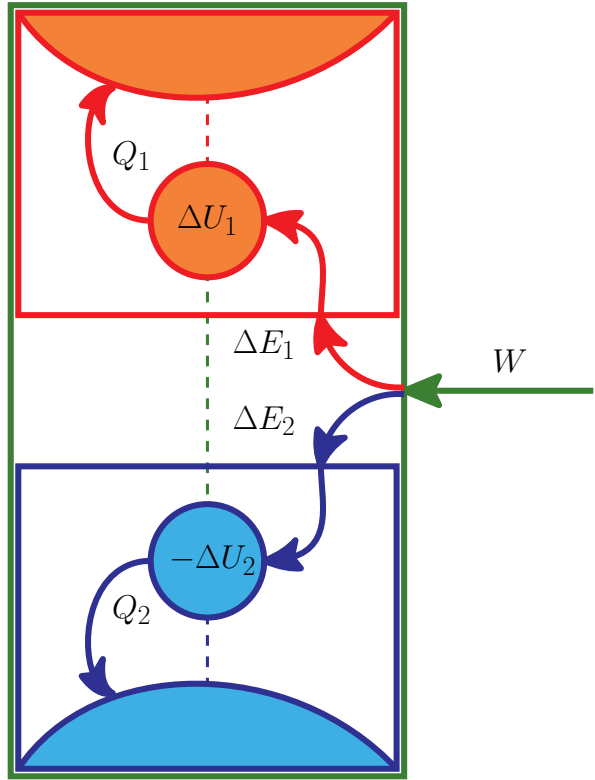}
		\caption{Scheme of a quantum thermal machine. An isolated system, (big green rectangle) is driven by an external time dependent field. The system is composed of two sub-systems (red and blue rectangles). Each subsystem is composed of a small quantum system (small circle) and a large system, namely a thermal reservoir (large circular section). The two small circles form the working substance, we shall call them working parts. The drive acts on the working substance, thus injecting work $W$ in the whole system. A part of it, $\Delta E_1$, is delivered to subsystem 1, via the working part 1. The rest, $\Delta E_2= W-\Delta E_1$, is delivered to subsystem 2, via the working part 2. Each working part retains a part of the delivered energy $\Delta U_i$, and dumps the rest $-Q_i=\Delta E_i-\Delta U_i$ into its reservoir. These energetic exchanges are possible due to possibly time-dependent couplings between the two working parts, and between each working part and its reservoir (dashed lines). At the beginning of the driving each subsystem is at thermal equilibrium with a given temperature $T_i$.}
		\label{fig:HeatEngineModel}
		\end{center}
\end{figure}

Consider a driven bi-partite system:
\begin{eqnarray}
H_S(t) = H_1 + H_2 +V(t) 
\end{eqnarray}
with factorized initial condition,
\begin{eqnarray}
\rho = \frac{e^{-\beta_1 H_1}}{Z_1} \otimes \frac{e^{-\beta_2 H_2}}{Z_2}
\label{eq:bi-Gibbs}
\end{eqnarray}
Without lack of generality we shall assume throughout this work $\beta_1 \leq \beta_2$, i.e., the first sub-system is assumed to be not colder than the second, at the initial time. Also we shall assume that at all times the 
Hamiltonian is time reversal symmetric  \cite{Messiah62Book}.
We further assume the compound system is thermally isolated 
and the driving is turned on at time $t=0$ and turned off at time $t=\tau$. At these two times simultaneous projective measurements of the energies of both sub-systems are performed, giving the results $E_{n_1}^1,E_{n_2}^2$ and  $E_{m_1}^1,E_{m_2}^2$, where $i=1,2$ and $E_k^i$ is the $k$-th eigenvalue of sub-system $i$. 
According to the quantum exchange fluctuation theorem \cite{Campisi10PRL105,Jarzynski04PRL92,Andrieux09NJP11}, it is
\begin{eqnarray}
\frac{P(\Delta E_1,\Delta E_2)}{\widetilde P(-\Delta E_1,-\Delta E_2)} = e^{\beta1 \Delta E_1+\beta_2 \Delta E_2}
\label{eq:XFR}
\end{eqnarray}
where $\Delta E_i=E_{m_i}^i-E_{n_i}^i$ is the observed energy change in sub-system $i$, $P(\Delta E_1,\Delta E_2)$ is the joint probability of observing $\Delta E_1$ and $\Delta E_2$, and $\widetilde P(-\Delta E_1,-\Delta E_2)$ is the joint probability of observing $-\Delta E_1$ and $-\Delta E_2$ when the reversed driving $V(\tau-t)$ is applied.
The driving $V(t)$ injects some amount of energy in the compound system:
\begin{eqnarray}
W = \Delta E_1 +\Delta E_2
\end{eqnarray}
which is in fact the work performed by the external driving source to drive the system. Part of this energy, $\Delta E_1$ goes into sub-system $1$, and part of it, $\Delta E_2$, goes into subsystem $2$.
Using the above equation to make the change of variable $\Delta E_2 \rightarrow W$, we obtain a fluctuation relation for the joint probability of work $W$ and $\Delta E_1$:
\begin{eqnarray}
\frac{P(\Delta E_1,W)}{\widetilde P(-\Delta E_1,-W)} = e^{(\beta_1-\beta_2)\Delta E_1 +\beta_2 W} 
\label{eq:FT-EW}
\end{eqnarray}
Multiplying by $\widetilde P(-\Delta E_1,-W)$ and integrating in $dW d\Delta E_1$, one obtains the integral form of the fluctuation relation
\begin{eqnarray}
\langle e^{(\beta_2-\beta_1)\Delta E_1 - \beta_2 W} \rangle = 1 
\end{eqnarray}
Using the Jensen's inequality as usual, one obtains from this
\begin{eqnarray}
\langle W \rangle \geq \eta_C \langle \Delta E_1 \rangle
\label{eq:W>etaDEltaE1}
\end{eqnarray}
where $\eta_C= 1 - \beta_1/\beta_2$ is Carnot's efficiency.
The above equations hold regardless of size of the two subsystems, as long as the assumptions
introduced are satisfied. 
In modelling a quantum thermal machine we shall consider each subsystem as composed of two parts, 
namely a heat reservoir and a small 
quantum system which constitutes a part of the working substance, see Fig. \ref{fig:HeatEngineModel}. We shall call the small quantum systems the working parts.
The driving is applied on the working substance. The received  work $W$, is shared
between sub-system 1 and 2 as $\Delta E_1 $ and $\Delta E_2$.
We allow for the possibility of a time dependence of the couplings between the reservoirs and the working parts, in which case we consider them as part of the time dependent part of $V(t)$ of the Hamiltonian.
This encompasses  continuous mode thermal machines, where the couplings between the working parts and their respective reservoirs are constant in time  and non-vanishing, and machines operating in discrete mode (via distinct strokes) where those couplings can be switched on and off during operation. The three level maser is an example of continuous mode engine while Carnot, Otto, Diesel engine etc. operate in discrete mode.

The average quantities $\langle \Delta E_1 \rangle$, $\langle \Delta E_2 \rangle $, $\langle W \rangle $ define the operation regime of the machine:
\begin{itemize}
\item  HEAT ENGINE: $\langle \Delta E_1 \rangle \leq 0$, $\langle \Delta E_2 \rangle \geq 0$, $\langle W \rangle \leq 0$
\item  REFRIGERATOR: $\langle \Delta E_1 \rangle \ge 0$, $\langle \Delta E_2 \rangle \le 0$, $\langle  W \rangle \ge 0$
\item  HEATER: $\langle \Delta E_1 \rangle \le 0$, $\langle \Delta E_2 \rangle \ge 0$, $\langle  W \rangle \ge 0$
\end{itemize}

When the thermal machine works as a heat engine, Eq. (\ref{eq:W>etaDEltaE1}) gives
\begin{eqnarray}
\frac{\langle W\rangle}{\langle \Delta E_1 \rangle} \leq \eta_C
\end{eqnarray}
This is the second law of thermodynamics as expressed for a heat engine.
Our derivation proves its universality based on the time-reversal symmetric unitary dynamics of the whole system, and the initial bi-Gibbsian preparation.
In a similar way, when the machine operates as refrigerator, one finds
\begin{eqnarray}
\frac{-\langle \Delta E_2 \rangle}{\langle W \rangle} \leq \frac{1}{\beta_2/\beta_1-1} = \eta_C^{R}
\end{eqnarray}

Before proceeding it is worth remarking that there is a freedom of arranging the 
position of the border between the two subsystems, i.e. to arrange the initial bi-Gibbsian equilibrium. In fact fluctuation relations for heat engines have been derived previously assuming the working substance is fully included in one of the two subsystem only, say subsystem 2 \cite{Campisi14JPA47,Sinitsyn11JPA4}. In that case
work $W$ is delivered to subsystem 2, which retains a part $\Delta E_2$  (shared between reservoir 2, $-Q_2$, and working substance $\Delta U_2$), an dumps the other other part $\Delta E_1=-Q_1$ directly into reservoir 1 as heat. That arrangement  is particularly useful for a machine working as heat engine, because Eq. (\ref{eq:W>etaDEltaE1}) would read $-\langle Q_1\rangle /\langle W\rangle \leq \eta_C$ as in standard thermodynamics books. 

Here we adopt instead the scheme in Fig. \ref{fig:HeatEngineModel} because we have in mind an implementation where the coupling of the working parts to the reservoirs is fixed  and cannot be manipulated, while one can turn the interaction between the working parts on and off. By keeping this coupling off, it is then straightforward to prepare each working part in thermal equilibrium with its own bath. This corresponds to the scenario depicted in Fig. \ref{fig:HeatEngineModel}.
With our arrangement the average energy $\langle \Delta E_1 \rangle$ can be identified with the heat $-\langle Q_1 \rangle$ only when the energy $\langle \Delta U_1 \rangle$ stored in the working part is null or negligible as compared to $\langle \Delta E_1 \rangle$ and $-\langle Q_1 \rangle$. This happens when the number of cycles is long and the working substance has a finite energy spectrum. Then $\langle \Delta U_1 \rangle$ remains bounded while $\langle \Delta E_1 \rangle$ and $-\langle Q_1 \rangle$ grow linearly in time. If the condition
 is met then $-\langle W\rangle /\langle Q_1\rangle  \simeq \langle W\rangle /\langle \Delta E_1\rangle \le \eta_C$.
Otherwise, if the condition is not met one can well have the ratio $-\langle Q_1 \rangle/\langle W \rangle$ be larger than $\eta_C$. This however does not have an impact on the second law of thermodynamics stating that a machine \emph{working in a cycle} (implying $\langle \Delta U_1 \rangle = \langle \Delta U_2\rangle $ = 0) cannot have an efficiency larger than Carnot's efficiency.

It is important to stress that the choice of borders and appropriate associate thermodynamic quantities is the key to obtaining exact transient fluctuation relations, like Eq. (\ref{eq:FT-EW}), that is fluctuation relations that hold regardless of the time duration of the process under investigation \cite{Campisi11RMP83,Campisi10PRL105,Gaspard13NJP15,Cuetara14PRE89}. In the long time limit, steady-state fluctuation relations hold which are independent of the border choice.

\section{Optimal two-qubit engine}
Our aim is to propose a minimal model of thermodynamic quantum engine which could be implemented and tested experimentally as a solid state quantum device. The simplest model one can think of is that of a single qubit coupled to two reservoirs at different temperatures. The qubit is driven by an external drive which changes its Hamiltonian in time for example by changing its energy spacing $H_\text{qubit}(t)=\omega(t) \sigma_z/2$. If one has the further ability to couple and decouple the qubit from the two reservoirs one can implement a 4-stroke engine, e.g. a Otto cycle. This can be realised, e.g., by interfacing the qubit to the thermal reservoirs by means of band-pass filters, as proposed in Ref. \cite{Niskanen07PRB76}. Here we focus instead on the case when the coupling to the reservoirs are fixed in time. In order to have a heat engine/refrigerator in continuous mode, a more complex working substance is necessary than a mere two level system. One needs a working substance that would be able to de-route the energy towards the wanted direction (from the hot bath to work source and cold bath for a heat engine; from the cold bath and the work source to the hot one for a refrigerator). For this reason we introduce a second qubit.
Qubit one is in contact with the first bath and qubit 2 is contact with bath 2, as in Fig. \ref{fig:HeatEngineModel}. A time dependent coupling $V(t)$ couples the two qubits for a time duration $[0,\tau]$.
The full Hamiltonian is:
\begin{eqnarray}
H(t) = H_{q,1} + H_{B,1} + H_{\text{int},1} + H_{q,2} + H_{B,2} + H_{\text{int},2}+ V(t) 
\end{eqnarray}
where $H_{B,i}$, $H_{\text{int},i}$, $i=1,2$, are the $i$-th bath Hamiltonian and its interaction with qubit $i$, respectively, and 
\begin{eqnarray}
H_{q,i} = \frac{\omega_i}{2} \sigma^z_i
\end{eqnarray}
is the the $i$-th qubit Hamiltonian. Here $\sigma^z_i$ denotes the $z$ Pauli sigma matrix of the $i$-th qubit.

To keep the discussion as simple and intuitive as possible we introduce a useful assumption, namely the coupling $V(t)$ is turned on for a time period $[0,\tau] $ that is very short compared to the relaxation time of each qubit to its own bath. The effect of the coupling $V(t)$ can accordingly be modelled by a unitary operator $U$ acting in the Hilbert space of the working substance, namely the two qubits. We shall call $U$  the gate operation.
The two qubits are initially each in thermal equilibrium with its own bath, i.e. their state is characterised by the density matrix
\begin{eqnarray}
\rho = \frac{e^{-\beta_1 H_{q,1}}}{Z_1} \otimes \frac{e^{-\beta_2 H_{q,2}}} {Z_2}
\label{eq:initial-state-WS}
\end{eqnarray}
with $Z_i = \Tr\,  e^{-\beta_i H_{q,i} }=2 \cosh (\beta_i \omega_i /2)$.
The average work injected into the working substance by applying the unitary $U$ is 
\begin{eqnarray}
\langle W \rangle &= \Tr (H_{q,1}+H_{q,2}) (U\rho U^\dagger- \rho)
\end{eqnarray}
and the energy taken by each sub-system is:
\begin{eqnarray}
\langle \Delta E_i \rangle &= \Tr H_{q,i} (U\rho U^\dagger- \rho)
\end{eqnarray}
If after the application of the gate $U$ each qubit is let interact with its respective reservoir for a sufficiently long time so as to reach the state of thermal equilibrium. During this thermalisation step they will give the heats  $-\langle Q_i \rangle = \langle \Delta E_i \rangle$ to the baths.

We are interested in the unitary that outputs the most work per cycle. Therefore we have searched for the unitary that maximises $\langle W \rangle $. We have pursued this task by parametrising a $4 \times 4$ unitary by means of $15$ angles as discussed in \cite{HedemannArXiv13035904} and performing a maximisation over the corresponding $15$ dimensional space. Numerics clearly indicates that maximum work output is achieved by means of the complex SWAP unitaries, reading in the $\{|+,+\rangle,|+,-\rangle,|-,+\rangle,|-,-\rangle\}$ basis:
\begin{eqnarray}
U=\left(\begin{array}{cccc}e^{i\phi_1} & 0 & 0 & 0 \\0 & 0 & e^{i\phi_2} & 0 \\0 & e^{i \phi_3} & 0 & 0 \\0 & 0 & 0 & e^{i \phi_4}\end{array}\right)\, .
\label{eq:Uswap}
\end{eqnarray}
With these $U$'s we find
\begin{eqnarray}
\label{eq:DeltaE1}\langle \Delta E_1 \rangle &= -\left( \frac{1}{1+e^{\beta_1 \omega_1}}-\frac{1}{1+e^{\beta_2 \omega_2}}
  \right)\omega_1\\
\label{eq:DeltaE2}  \langle \Delta E_2 \rangle &= 
\left( \frac{1}{1+e^{\beta_1 \omega_1}}-\frac{1}{1+e^{\beta_2 \omega_2}}
  \right)\omega_2\\
\label{eq:W}    \langle W \rangle &= 
\left( \frac{1}{1+e^{\beta_1 \omega_1}}-\frac{1}{1+e^{\beta_2 \omega_2}}
  \right)(\omega_2-\omega_1)
\end{eqnarray}
In the following we fix the gate to be any complex swap gate in Eq. (\ref{eq:Uswap}). Quite remarkably, the same unitaries also maximise the heat engine efficiency. \subsection{Operation}
The operation of the swap-machine is dictated by the relative signs of $\langle \Delta E_1 \rangle$,
$\langle \Delta E_2 \rangle$, $\langle W \rangle$. With $\beta_1 \leq \beta_2$ the conditions for each mode of operation are:
\begin{itemize}
\item  HEAT ENGINE: $\frac{\beta_1}{\beta_2}<\frac{\omega_2}{\omega_1}<1$
\item REFRIGERATOR :  $0<\frac{\omega_2}{\omega_1}<\frac{\beta_1}{\beta_2}$
\item  HEATER:  $1<\frac{\omega_2}{\omega_1}$
\end{itemize}
\begin{figure}
		\begin{center}
		\includegraphics[width=\linewidth]{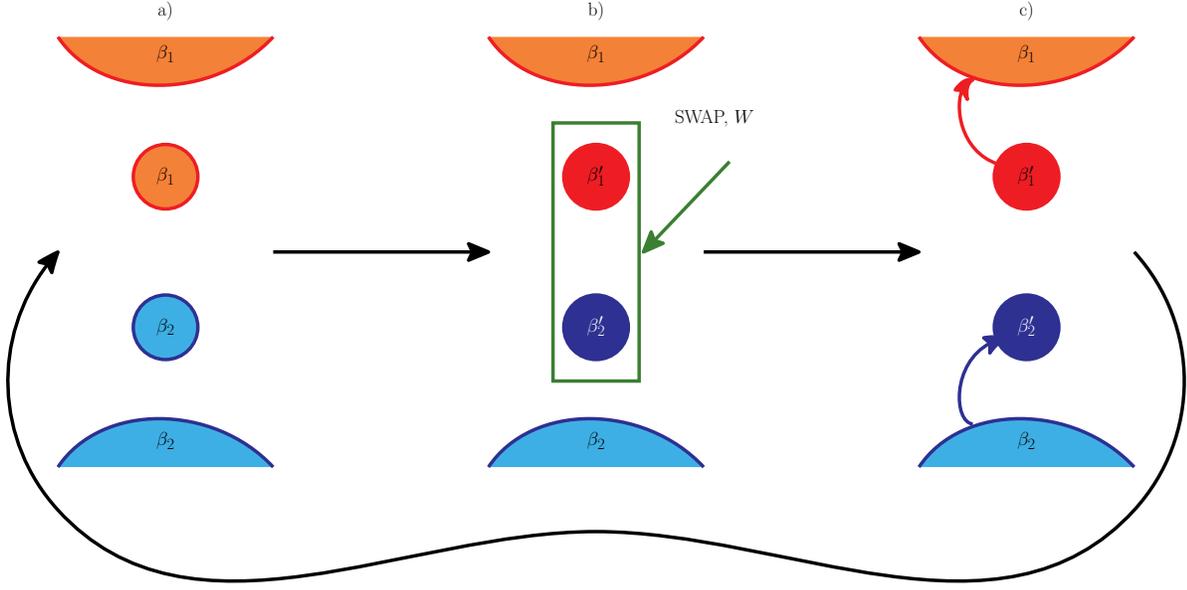}
		\caption{Illustration of the functioning of the SWAP engine in the refrigerator mode. a) Each qubit is in thermal equilibrium with its respective bath. b) The instantaneous SWAP gate is applied resulting in the injected work $W$. The hot qubit becomes hotter while the cold qubit becomes colder. c) Qubit 1 cedes heat to the hot bath. Qubit 2 withdraws heat from the cold bath. The initial equilibrium a) is re-established. In heat engine mode the swap cools the hot qubit and heats the cold qubit, while outputting work. The sign of the heat and work arrows gets inverted accordingly.}
		\label{fig:Cartoon}
		\end{center}
\end{figure}
The explanation of the above conditions is as follows. After the SWAP-gate operation is performed the two qubits are in the states
\begin{eqnarray}
\rho_1' & \propto e^{-\beta_2 \omega_2 \sigma^1_z/2} =e^{-\beta'_1  H_{q,1}} \\
\rho_2' &\propto e^{-\beta_1 \omega_1 \sigma^2_z/2} = e^{-\beta'_2  H_{q,2}}
\end{eqnarray}
where 
\begin{eqnarray}
\beta'_1= \beta_2 \omega_2/\omega_1  \\
 \beta'_2  = \beta_1 \omega_1/\omega_2 
\end{eqnarray}
If $\omega_2/\omega_1 < \beta_1/\beta_2$ then $\beta_1'<\beta_1$ and $\beta_2' > \beta _2$, hence the cold qubit cools down and the hot qubit heats up $\langle \Delta E_1 \rangle>0$, $\langle \Delta E_2 \rangle<0$. Also, since $\omega_2/\omega_1 < \beta_1/\beta_2<1$, then $\langle W \rangle >0$. Hence we have the fridge operation.
If $\omega_2/\omega_1 > \beta_1/\beta_2$ then $\beta_1'> \beta_1$ and $\beta_2' < \beta _2$, hence the hot qubit cools down and the and the cold qubit heats up $\langle \Delta E_1 \rangle<0$, $\langle \Delta E_2 \rangle>0$. In this case, depending on the relative size of $\omega_1$ and $\omega_2$ we will have either heat engine or heater. Let $u(x)= \Tr \sigma_z e^{-x \sigma_z}/\Tr e^{-x \sigma_z}$. Then $\langle \Delta E_1 \rangle = \omega_1 [ u (\beta_2 \omega_2)- u (\beta_1 \omega_1) ]$ and 
$\langle \Delta E_2 \rangle = \omega_2 [ u (\beta_1 \omega_1)- u (\beta_2 \omega_2) ] = -(\omega_2 /\omega_1) \langle \Delta E_1 \rangle $. Accordingly 
$\langle W \rangle = (1-\omega_2 /\omega_1 )\langle \Delta E_1 \rangle$. If $\omega_2 /\omega_1<1$, then $\langle W \rangle <0$ and we have the heat engine. Otherwise dud engine. 
Fig. \ref{fig:Cartoon} shows a cartoon of the operation of the machine in the refrigerator mode.

\subsection{Efficiency}
For the heat engine operation, it is $0<\frac{\omega_2}{\omega_1}<\frac{\beta_1}{\beta_2}$, hence:
\begin{eqnarray}
\eta =\frac{-\langle W\rangle}{-\langle \Delta E_1\rangle} = 1-\frac{\omega_2}{\omega_1} \leq 1-\frac{\beta_1}{\beta_2} = \eta_C
\end{eqnarray}
For a fridge
\begin{eqnarray}
\eta^F=\frac{-\langle \Delta E_2\rangle}{\langle W\rangle} = \frac{\omega_2}{\omega_1-\omega_2} = \frac{1}{\omega_1/\omega_2-1} \leq \frac{1}{\beta_1/\beta_2-1}= \eta_C^F
\end{eqnarray}
because for the fridge $\frac{\beta_1}{\beta_2}<\frac{\omega_2}{\omega_1}<1$. 

Note that in order for the engine to function the two qubits must have different energy spacings $\omega_i$, otherwise the work intake (output), will be exactly null. Not also that the efficiency depends only on the ratio $\omega_2/\omega_1$ and not on the temperatures $\beta_1,\beta_2$. This is a peculiar feature of the swap unitary. In the following we shall focus on heat engine operation.

\subsection{Efficiency at maximum power}
\begin{figure}[t]
		\begin{center}
		\includegraphics[width=\linewidth]{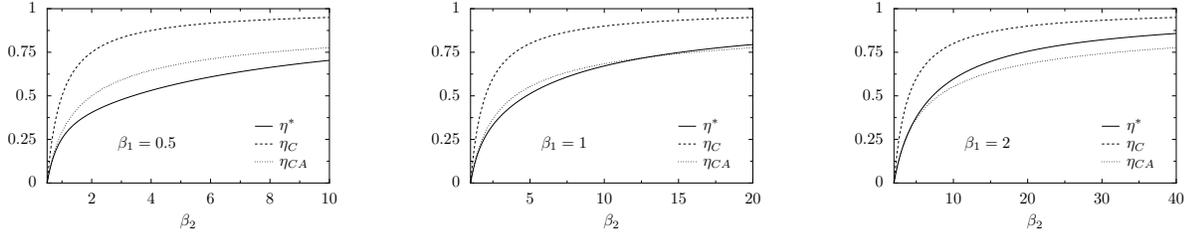}
		\caption{Efficiency at maximum power as function of $\beta_1$ for various values of $\beta_1$.}
		\label{fig:etaMP}
		\end{center}
\end{figure}
Given the two temperatures $T_1,T_2$ the maximal efficiency, i.e., Carnot's efficiency is reached when $\omega_2/\omega_1 \rightarrow \beta_1/\beta_2$. In this regime however, the work tends to zero, see Eq. (\ref{eq:W}).
It is interesting that here the power at Carnot efficiency is zero, as with standard strokes engines, but not because of slow operation. 

On the other hand, given the two temperatures $T_1, T_2$ one can find the value of $\omega_1$ and $\omega_2$ for which the 
power output, $-\langle W \rangle$, is maximum. This can be achieved by maximising the work output in Eq. (\ref{eq:W}).
The maximum depends indeed only on the ratio $\Omega = \omega_2/\omega_1$.
This can be best seen by setting $\omega_1$ as the unit of energy, so that $\omega_1=1$, and all energies are measured as multiples of $\omega_1$. With these units 
\begin{eqnarray}
\langle W \rangle =  \left( \frac{1}{1+e^{\beta_1}}-\frac{1}{1+e^{\beta_2 \Omega}}\right) (\Omega-1) 
\end{eqnarray} 
We denote the value of $\Omega$ for which $-\langle W \rangle$ is maximum at given $\beta_1,\beta_2$ as $\Omega^*(\beta_1,\beta_2)$.
The corresponding efficiency, namely the efficiency at maximum power is:
\begin{eqnarray}
\eta^*(\beta_1,\beta_2) = 1- \Omega^*(\beta_1,\beta_2)
\end{eqnarray}
For example, the value of $\Omega^*$  is  $\Omega^*=0.83$ for $k_B T_1=3/2, k_B T_2=1$ (in units of $\omega_1$ as explained above). The corresponding efficiency at maximum power is $\eta^* \simeq 0.17$.

Figure \ref{fig:etaMP} shows the maximum power efficiency $\eta^*(\beta_1,\beta_2)$ as a function of $\beta_2$ for various fixed values of $\beta_1$. The figure also report the corresponding Carnot efficiency and the Curzon-Albhorn efficiency \cite{Curzon75AJP43}
\begin{eqnarray}
\eta_\text{CA} = 1 - \sqrt{\frac{\beta_1}{\beta_2}} 
\end{eqnarray}
The figure shows that the the maximum power efficiency can be both larger and smaller than the Curzon-Albhorn efficiency. 
However for sufficiently low $\beta_1$ (hotter hot reservoir), $\eta^* < \eta_\text{CA}$, while, for sufficiently high $\beta_1$, (colder hot reservoir), $\eta^* > \eta_\text{CA}$, that is at very low temperature the efficiency at maximum power is above the Curzon-Albhorn efficiency.

\begin{figure}[t]
		\begin{center}
		\includegraphics[width=.5\linewidth]{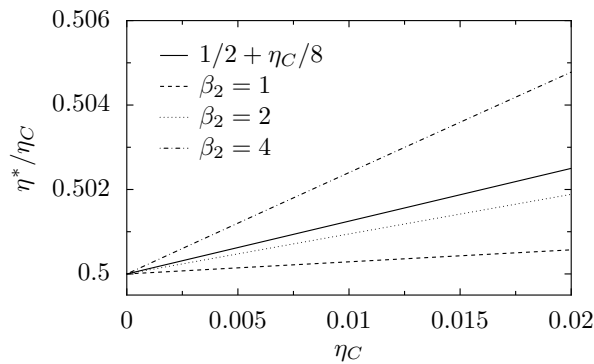}
		\caption{Low $\eta_C$ behaviour of $\eta^*$. The plot shows that $\eta^*$ behaves as in Eq. (\ref{eq:lowEtaC}). }
		\label{fig:etaMPLowEtaC}
		\end{center}
\end{figure}

We have performed an analysis of the maximum power efficiency $\eta^*$ for $\beta_1 \simeq \beta_2$, i.e., in the low $\eta_C$ limit.  In accordance to linear response theory we expect the linear coefficient of the expansion to match the value $1/2$ \cite{VandenBroeck05PRL95}. Since our engine is not a thermoelectric engine (work is provided by time-dependent pulses, rather than by a DC electric potential difference), and does not have the left-right symmetry (in order for it to output some work the energy spacings of the two qubits, $\omega_1$ and $\omega_2$, should be different), we do not expect that the value $1/8$ for the quadratic coefficient, predicted in those cases Ref. \cite{Esposito09PRL102}, to be obeyed. The results of the low $\eta_C$ analysis, reported in Fig. \ref{fig:etaMPLowEtaC}, corroborate these expectations. The figure presents plots of $\eta^*/\eta_C$ for various values of $\beta_2$ as function of $\eta_C$. The plots clearly show that 
\begin{eqnarray}
  \eta^* \simeq \frac{\eta_c}{2} + f(\beta_2){\eta_c^2} + O(\eta_C^3)\, ,
  \label{eq:lowEtaC}
\end{eqnarray}
that is, the linear coefficient $1/2$ is obeyed whole the quadratic coefficient is a function $f(\beta_2)$ whose value may differ from $1/8$.

\section{Modelling: Quantum jumps}
The above analysis based on the simplified assumption of unitary gate followed by thermalisation, allowed us to make predictions about the average work and heats that go in the two reservoirs. It does not suffice however for the full stochastic characterisation of the engine. In order to achieve that we need to model the dynamics of the thermalization. We assume then that the effect of the thermal environment on each qubit can be modelled by means of a master equation of Lindblad form.
\begin{equation}
\dot \rho_i = -i [H_{q,i},\rho] + \mathcal L_i \rho 
\end{equation}
where 
\begin{eqnarray}
\mathcal L_i \rho &= \gamma (n_i+1) D[\sigma_i] \rho + \gamma n_i D[\sigma^\dagger_i]\rho_i,  \qquad	i = 1,2\\
n_i &= \frac{1}{e^{\beta_i \omega_i}-1}   \qquad	\\
D[c]\rho &= c\rho c^\dagger -\frac{1}{2} c^\dagger c \rho- \frac{1}{2}\rho c^\dagger c
\end{eqnarray}
and $\sigma_i = \sigma_i^x + i \sigma_i^y$ is the annihilation operator for the qubit $i$, and $\sigma_i$, its adjoint, is the creation operator.

To obtain the statistics of energy exchanges with the bath during the thermalisation step, we proceed to un-ravel the master equation \cite{BreuerPetruccioneBOOK}, as proposed in \cite{Horowitz12PRE85} and \cite{Hekking13PRL111}. This results in a stochastic differential equation in the Hilbert space of each qubit:
\begin{eqnarray}
d |\psi_i \rangle = &-&i G_i(|\psi_i \rangle) dt \nonumber \\
&+& \left(\frac{\sigma_i |\psi_i \rangle }{|| \sigma_i |\psi_i \rangle ||} - |\psi_i \rangle \right) dN_i^+ 
+ \left(\frac{\sigma_i^\dagger |\psi_i \rangle }{|| \sigma_i^\dagger |\psi_i\rangle ||} - |\psi_i \rangle \right) dN_i^-
\end{eqnarray}
The deterministic part is given by:
\begin{eqnarray}
G_i(|\psi_i \rangle) &= H_{q,i}^\text{eff} |\psi_i\rangle 
+ \frac{i}{2} \gamma (n_i+1) || \sigma_i |\psi_i \rangle ||^2 |\psi_i \rangle 
+ \frac{i}{2} \gamma (n_i) || \sigma_i^\dagger |\psi_i \rangle ||^2 |\psi_i \rangle \\
H_{q,i}^\text{eff} &= H_{q,i} -\frac{i}{2}\gamma  (n_i+1)\sigma_i^\dagger \sigma_i -\frac{i}{2}\gamma  n_i  \sigma_i \sigma_i^\dagger \nonumber \\
&= H_{q,i} -\frac{i}{2}\gamma n_i -i\gamma \sigma_i^\dagger \sigma_i
\end{eqnarray}
while the stochastic Poisson increments have the ensemble expectations
\begin{eqnarray}
E(dN_i^+) &= \gamma  (n_i+1) || \sigma_i |\psi_i \rangle ||^2 dt \\
E(dN_i^-) &= \gamma  (n_i) || \sigma_i^\dagger |\psi_i \rangle ||^2 dt 
\end{eqnarray}
The stochastic equations can be solved by means of the Monte Carlo Wave Function (MCWV) method \cite{Molmer93JOSAB}. In the present case of an undriven single qubit they result in a dichotomic Poisson process governed by the two rates:
\begin{eqnarray}
\Gamma_i^-= \gamma n_i \\
\Gamma_i^+= \gamma (n_i+1)
\end{eqnarray}
depending on whether the qubit is in the down state $|-\rangle$ or up state $|+\rangle$. Note that these rates are detailed balanced:
\begin{eqnarray}
\frac{\Gamma_i^-}{\Gamma_i^+}= e^{-\beta_i \omega_i}
\end{eqnarray}
\begin{figure}[t]
		\begin{center}
		\includegraphics[width=\linewidth]{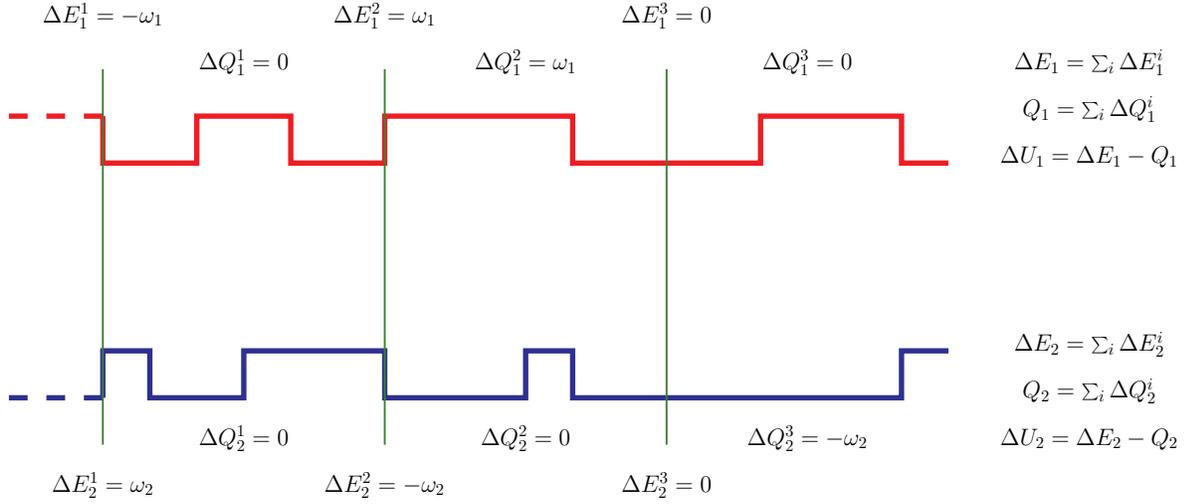}
		\caption{Typical quantum trajectory of the working substance of the SWAP engine. Top trajectory is for qubit 1. Bottom trajectory is for qubit 2. Green vertical  lines indicate times when the instantaneous SWAP gate is applied. Transitions occurring at these time are due to the work done by the external work source. Transitions occurring between the SWAP pulses signal heat exchanges with the heat reservoirs.}
		\label{fig:trajectory}
		\end{center}
		\label{fig:Fig2}
\end{figure}
Ref. \cite{Horowitz12PRE85} has studied the fluctuation relations for such quantum trajectories but for systems being in contact with a single thermal reservoir. Here our working substance, the two qubits, is in contact with two distinct reservoires. The analysis performed in \cite{Horowitz12PRE85} can however be extended to multiple reservoirs. It results in the following fluctuation relation for the probability of a given quantum trajectory $\gamma$:
\begin{eqnarray}
\frac{P[\gamma]}{P[\widetilde \gamma]} = \exp\left( \beta_1Q_1[\gamma]+ \beta_2 Q_2[\gamma] + \ln \frac{p_a}{p_b}\right)
\label{eq:trajectoryFR}
\end{eqnarray}
where $\widetilde \gamma$ is the time reverse of $\gamma$. We remark that in the case of multiple reservoirs $\gamma$ is  not only specified by the temporal evolution of the state of the central system (the working substance in our case), call it $\chi_t$, but also by the succession $i_n$ indicating which bath (labelled by $i$) was responsible for each of the $N$ jumps (labelled by $n$), $\gamma= (\{\chi_t\}, \{i_n\})$ Accordingly the time reversed trajectory results by requiring that the temporal evolution of the central system state is inverted and if the $n$-th jump of the forward trajectory $\gamma$ was caused by the $i$-th bath, so was the last $n$-th jump of the backward trajectory $\widetilde \gamma$. That is $\widetilde \gamma_t = (\{\chi_{\mathcal T -t}\}, \{i_{N-n}\}) $. In our case the trajectory $\gamma$ has two components $\gamma_t= (\gamma_{1,t},\gamma_{2,t})$, each specifying the temporal evolution of the state of each qubit. No extra indexes are necessary because all jumps in $\gamma_{1,t}$ are caused by reservoir $1$ and all jumps in $\gamma_{2,t}$ are caused by reservoir $2$. Accordingly 
$\widetilde \gamma_t = (\gamma_{1,\mathcal T -t},\gamma_{2,\mathcal T -t})$. 
 The symbol $Q_i[\gamma]$, means the heat ceded to the $i$-th reservoir during the realisation of $\gamma$. Specifically $Q_i[\gamma]=\int_0^{\mathcal T} \omega_i (dN_{i,s}^- -dN_{i,s}^+)d s$. Obviously in our case $Q_i$ is a functional of $\gamma_i$ only. In Eq. (\ref{eq:trajectoryFR}) $a,b$ denote the initial and final state of the trajectory $\gamma$, i.e. $\gamma_0=a$, $\gamma_{\mathcal T}=b$, and $p_{a,b}$ are the respective probability that these states are observed. With our choice (\ref{eq:initial-state-WS}) it is $p_x  =\exp(-\beta_1 U_1^x -\beta_2 U_2^x)/(Z_1Z_2)$, $x=a,b$.
Writing $\Delta U_i = U_i^b-U_i^a$, and using $\Delta U_i = \Delta E_i[\gamma] - Q_i[\gamma]$ we obtain
\begin{eqnarray}
\frac{P[\gamma]}{P[\widetilde \gamma]} = \exp\left( \beta_1\Delta E_1[\gamma]+ \beta_2 \Delta E_2[\gamma]\right)
\end{eqnarray}
Multiplying by $P[\widetilde \gamma] \delta(\Delta E_1- \Delta E_1 [\gamma])\delta(\Delta E_2- \Delta E_2 [\gamma])$ and performing a path integral over all trajectories $\gamma$ one recovers Eq. (\ref{eq:XFR}). Accordingly all subsequent relations in Sec. \ref{sec:HEFR} are obeyed within our quantum jump modelling.

\section{Stochastic thermodynamics of the SWAP  engine}
\begin{figure}[t]
		\begin{center}
		\includegraphics[width=\linewidth]{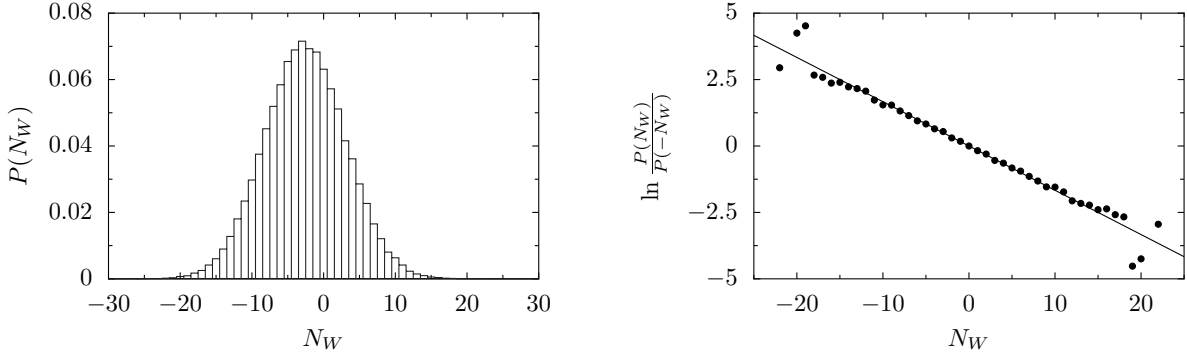}
		\caption{Left panel: Probability $P(N_W)$ of work quanta $N_W$ given off by the work source. Right Panel: Corresponding logarithmic fluctuation ratio $\ln [P(N_W)/P(-N_W)]$. Dots: numerics. Solid line: theoretical line $(\beta_1 \omega_1-\beta_2 \omega_2)N_W$. Discerpancy at large $N_W$ is ascribed to bad corresponding statistics. The histogram $P(N_W)$ was constructed from a sample of $10^6$ trajectories. Here $k_BT_1=1.5$,$k_BT_2=1$, $\omega_1=1$, $\omega_2=5/6$, corresponding to heat engine operation. The time between swaps $\tau_2\simeq 0.65 $ was about half the relaxation time $\tau_\text{relax}$ and $N=100$ swap gates were applied.}
		\label{fig:workPDF}
		\end{center}
\end{figure}
We operate the machine in the following manner. At time $t=0$ we pick up a state randomly from the initial bi-Gibbsian distribution, Eq. (\ref{eq:initial-state-WS}). We apply the complex SWAP gate, Eq. (\ref{eq:Uswap}), and generate the stochastic dynamics of each qubit using the MCWF method until time $\tau_2$, when we apply the complex SWAP again, and let evolve stochastically until time $2\tau_2$, and so on for a total duration $\mathcal T = N\tau_2$. Our assumption is that the swap gate is much faster than the stochastic evolution time: $\tau \ll \tau_2$.
Figure \ref{fig:Fig2}  shows a sketch of the resulting quantum trajectories of the two qubits, along with the energetic exchanges the various jumps signal.

The first important observation from Fig. \ref{fig:Fig2} is that any time the energy $\Delta E_1$ is given to subsystem 1, accordingly the energy $\Delta E_2 = -(\omega_2/\omega_1)\Delta E_1$ is taken from subsystem 2.
This implies that all trajectories have the same efficiency $\eta =  W /\Delta E_1 = (\Delta E_1+ \Delta E_2)/\Delta E_1 = 1-\omega_2/\omega_1 = \eta $. In other words there are no efficiency fluctuations. This is because the gate swaps the eigenstates of the double qubit without creating superpositions thereof. Hence each swap pulse $k$ deterministically and univocally  results in well defined values of $\Delta E_{1,2}^k$ depending on the state of each qubit before its application. A generic unitary will typically create a superposition of the eigenstates, which can collapse either in the up state or down state of each qubit with according probability. The value of   $\Delta E_{1,2}^k$ would be accordingly not uniquely defined by the state before a generic gate.

The constraint $\Delta E_2 = -(\omega_2/\omega_1)\Delta E_1$ allows to express the heat engine fluctuation relation (\ref{eq:FT-EW}) as a relation for a single variable, say $W$. Since  $W$ is an integer multiple $N_W$ of $\omega =\omega_1-\omega_2$, the fluctuation relation can be conveniently expressed as a relation for the probability $P(N_W)$ that $N_W$ of work quanta are given off by the work source. We obtain then
\begin{eqnarray}
\frac{P(N_W)}{P(-N_W)}= e^{(\beta_1 \omega_1-\beta_2 \omega_2)N_W}
\label{eq:HEFR-NW}
\end{eqnarray}

Figure \ref{fig:workPDF} shows $P(N_W)$ and the corresponding logarithmic ratio $\log P(N_W)/P(-N_W)$ for one simulation of our engine. In an experimental realisation the probability $P(N_W)$, can be constructed by recording the number and sign of the swaps occurred during each of many realisations in just one of the two qubits.

\begin{figure}
		\begin{center}
		\includegraphics[width=\linewidth]{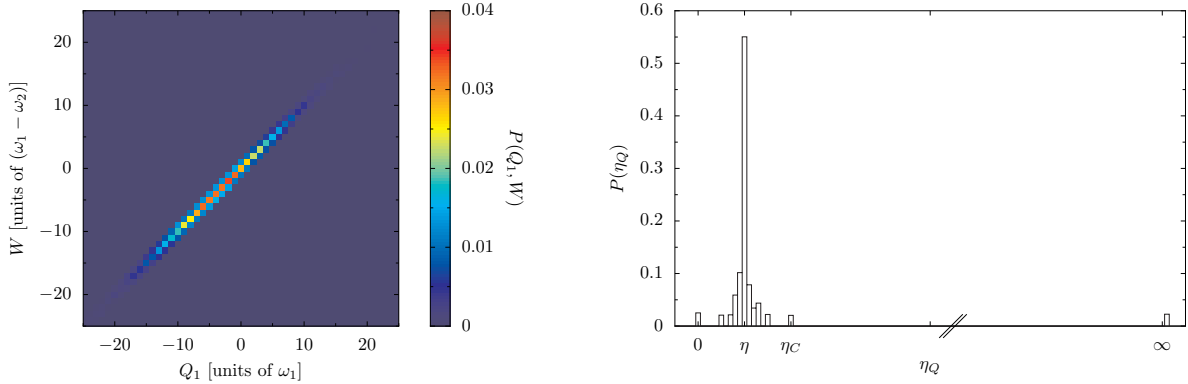}
		\caption{Left panel: joint distribution $p(Q_1,W)$. Right panel the corresponding $Q$-efficiency distribution $P(\eta_Q)$.
The plots are for $k_B T_1=3/2, k_B T_2=1,\omega_1=1, \omega_2=5/6;$ corresponding to the heat engine regime. The number of applied pulses is $N=100$. The sample consists of $10^6$  trajectories. Carnot efficiency is $\eta_C= 1-2/3 = 1/3$. $\Delta E$-efficiency is $\eta= 1- 5/6= 1/6 = \eta_C/2$.}
		\label{fig:JointPQW}
		\end{center}
\end{figure}

In Fig. \ref{fig:JointPQW}, left panel, we report a plot of the joint probability distribution of heat and work $P(Q_1,W)$.
Note how $W/(\omega_1-\omega_2)$ differs from $Q_1/\omega_1$ at most by one unit. This is because $\Delta E_1/\omega_1$ differs from $Q_1/\omega_1$ at most by one unit, i.e. one quantum of energy stored in qubit one as $\Delta U_1$.
Because $Q_1$ is not exactly equal to $\Delta E_1$, the heat-efficiency $\eta_Q = -W/Q_1$ has some fluctuations, in contrast to the $\Delta E$-efficiency $\eta=-W/\Delta E_1 $. The statistics of $\eta_Q$ corresponding to the plot in Fig. \ref{fig:JointPQW} is reported in the right panel of Fig. \ref{fig:JointPQW}.

Note the very pronounced peak at $\eta$. Note also a second peak $\eta^C$. We observe that there is a finite probability that $\eta_Q$ is infinite. Because of the peak at infinity the quantity $\langle \eta_Q \rangle$ is not well defined. As the number of cycles increases the spot in Fig 5 drifts and diffuses in the diagonal direction, but not in the transverse direction. A consequence of this is that the peak at $\eta$ in the efficiency probability increases while all other peaks decay. That is for large operation time the probability of $\eta_Q$ coincides with the probability of $\eta$ as expected. \footnote{According to large deviation theory, all peaks but the most likely fall with an exponential rate, the largest of which is for the Carnot efficiency \cite{Verley14NATCOMM5}.}

\subsection{Increasing the power}
\begin{figure}[t]
		\begin{center}
		\includegraphics[width=.5	\linewidth]{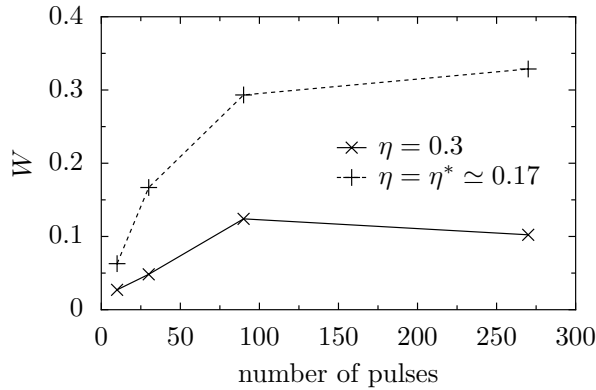}
		\caption{Efficiency and work output as a function of number of applied pulses for a fixed operation time. Here $k_B T_1=3/2, k_B T_2=1$ (ini units of $\omega_1$). The operation time of the machine is fixed and equal to $T_\text{op}=30 \tau_\text{relax}$ ($\tau_\text{relax}$ is defined here as the longest among the thermal relaxation times, i.e., $\tau_\text{relax}=\gamma^{-1}\max[e^{\beta_1\omega_1}-1,e^{\beta_2\omega_2}-1  ]$). At $N=10$ cycles, the engine has plenty of time to relax to equilibrium, because each pulse is followed by a rest time of $3 \tau_\text{relax}$. By increasing the pulse frequency one can greatly enhance the work output. Solid line, $\omega_1=1, \omega_2=0,7$, corresponding to efficiency $\eta=0.3$.
		Dashed line:  $\omega_1=1, \omega_2\simeq 0.83$ corresponding to efficiency at max power $\eta^*=  \simeq 0.17$. }
		\label{fig:increasingPower}
		\end{center}
\end{figure}

As discussed above the swap heat engine works at the efficiency $\eta=1- \omega_2/\omega_1$ regardless of the power output.
This is a great advantage over traditional engines because increasing the power has no cost in terms of reducing the
efficiency for our engine. The power of our engine can be increased simply by increasing the swap-pulses frequency.
This is illustrated in Fig. \ref{fig:increasingPower}.
Figure \ref{fig:increasingPower} suggests that the power output saturates at a regime value as the pulse frequency increases. The saturation value gives the maximum power for the given $T_1, T_2,\omega_1, \omega_2$. It is important to stress that for too frequent swap pulses, namely when their temporal separation $\tau_2$ is of the same order as the temporal duration $\tau$ of the swap pulse, our simplifying assumption (namely that dynamics can be modelled separately by a unitary followed by the stochastic relaxation) does not hold any more. In a real experiment the power output is expected to decay in the range of highly frequent pulses.

\section{Solid state implementation and measurement scheme}
For our implementation proposal we follow the scheme presented in Ref. \cite{Pekola14NJP15}, with the necessary
modifications and extensions. Ref. \cite{Pekola14NJP15} presents an experimental scheme where a single Cooper-Pair-Box
(CPB), namely a qubit, is coupled to a resistor at some temperature $T$. The resistor comprises an electronic system coupled to a phononic one. When a photon is emitted (absorbed) into the resistor, the fast electronic system responds by heating up (cooling down) abruptly, and then relaxing to the thermal equilibrium set by the phononic substrate. A nano calorimeter can then be used to monitor the temperature of the electronic system, in order to detect absorbed/emitted photons. As reported in \cite{Gasparinetti14arXiv}, sufficienty fast and sensitive calorimeters for this pourpose are currently under development and should be soon available.

\begin{figure}[t]
		\begin{center}
		\includegraphics[width=.9\linewidth]{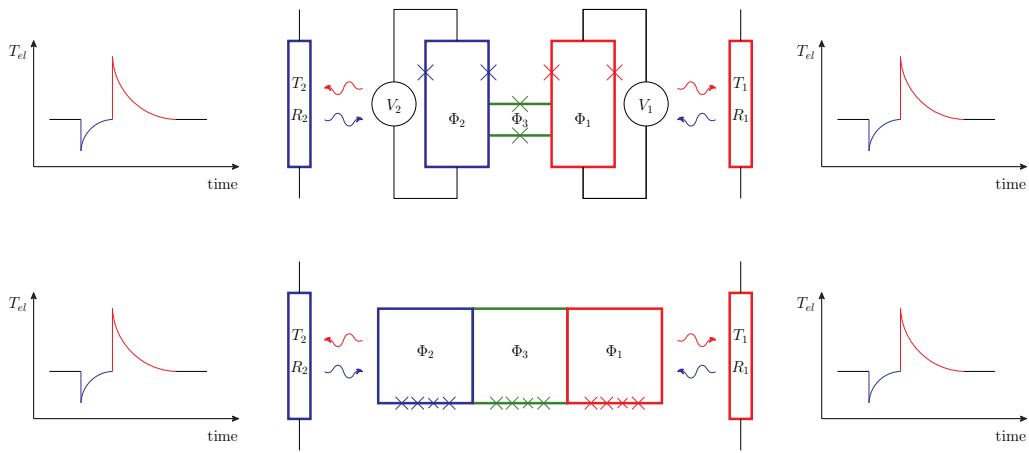}
		\caption{Scheme of two possible experimental set-up. Top Panel: Two Cooper Pair Boxes are coupled by means of two Josephson junctions in parallel as from Ref. \cite{Echternach01QIC1}. Bottom panel: Two flux qubits are coupled via a third flux qubit as from Ref. \cite{Niskanen07SCIENCE316}. Both setups can be used to realise the $i$SWAP gate, Eq. (\ref{eq:iSWAP}). The two qubits exchange photons each with a different resistor kept at a given temperature. Two on-chip fast calorimeters detect single photon emission/absorption in each resistor.}
		\label{fig:scheme}
		\end{center}
\end{figure}

In order to realise the SWAP engine, two such CPB + resistor systems should be realised on the same microchip, which does not seem to pose any particular difficulty. Each resistor is then monitored by an on chip calorimeter of the type in Ref. \cite{Gasparinetti14arXiv}. At variance with the set-up proposed in Ref. \cite{Pekola14NJP15} here the two CPBs have fixed energy gaps, hence they exchange photons of well defined energy $\hbar \omega_i$. This simplifies the measurement, because each calorimeter needs not measure the energy of the absorbed/emitted photon, but should just detect that a photon has been absorbed/emitted.
The gate operation can be implemented by coupling the two CPB using two tunnel junctions connected in parallel as described in 
\cite{Echternach01QIC1}. This allows for the implementation of the $i$SWAP gate, namely the complex SWAP gate

\begin{eqnarray}
U=\left(\begin{array}{cccc}1 & 0 & 0 & 0 \\0 & 0 & i & 0 \\0 & i & 0 & 0 \\0 & 0 & 0 & 1\end{array}\right)\, .
\label{eq:iSWAP}
\end{eqnarray}Fig. \ref{fig:scheme}, top panel, shows the scheme of the implementation.

An alternative implementation uses flux qubits operating at the optimal point, which have much longer coherence and relaxation times as compared to CPBs
\cite{Chiorescu03SCIENCE299,Mooij99SCIENCE285}. The switchable coupling is realised by means of a third qubit sandwiched between the two qubits as demonstrated in Ref. \cite{Niskanen07SCIENCE316}. $i$SWAP gate can be realised by means of microwave driving for a targeted time duration 
 \cite{Niskanen07SCIENCE316}. As demonstrated in Ref. \cite{Niskanen07SCIENCE316}, the minimum time for a universal gate is, in that set-up, about 22ns, while decoherence and relaxation times are at least 0.2 $\mu$s. This is in agreement with our assumption of fast gate operation, as compared to thermal relaxation.
 Fig. \ref{fig:scheme}, bottom panel, shows this alternative implementation.

By means of the calorimetric measurement one can experimentally access the quantum trajectories of the type shown in Fig. \ref{fig:Fig2}.
This is achieved in the following way.
The calorimeters can only detect the heat quanta $\Delta Q_k^i$ ceded to the resistors $i=1,2$. If two consecutive emissions (absorptions) are observed to occur in the same bath $i$, it means that meanwhile a quantum of energy $\Delta E^i_k =+(-) \omega_i$ has been given to (taken from) the $i^\text{th}$ qubit by the work source. 
Summing up all the $\Delta E^i_k$ one obtains the total energy given to each subsystem, $\Delta E_1,\Delta E_2$ and the work $W= \Delta E_1+\Delta E_2$. Having $Q_1,Q_2,\Delta E_1,\Delta E_2,W$ one can address the full statistics of energetic exchanges of the engine,
and accordingly can check the validity of the fluctuation relations (\ref{eq:XFR},\ref{eq:FT-EW}). Note that the measurement apparatus can also be employed to check the coincidence of swap-induced jumps in the two qubits, thus quantifying the goodness of the swap operation.

The employment of flux qubit for implementing on chip coolers have been discussed also in Ref. \cite{Niskanen07PRB76}. In the work of Ref. \cite{Niskanen07PRB76}  the working substance is a single flux qubit which is alternatively coupled and decoupled from the two baths. This is attained by embedding the two bath-resistors each in a LCR circuit, acting as band-pass filters centred at different frequencies $\omega_1,\omega_2$. As the qubit level spacing is switched between these two values the qubit interacts primarily with one resistor or the other so as to realise a Otto cycle, where interactions with the cold and hot bath occur in alternation and are separated by slow, adiabatic drives. This realises the same average heat and work exchanges as the present engine, hence same efficiency, with the difference that the present engine works in continuous mode. Heat exchanges with the two baths occur here simultaneously, and no adiabatic drive is employed. This makes it more promising in regard to the delivered power.

\section{Conclusions}
Based on a previous work \cite{Campisi14JPA47}, we have here presented a detailed discussion of fluctuation relations for heat and work in quantum heat engines. These fluctuations are illustrated by means of an optimal two-qubit engine working in continuous mode. We studied its full stochastic energetic exchanges including the statistics of its efficiency. At the average level this engine realises the same thermodynamics as the single qubit Otto engine of Ref. \cite{Niskanen07SCIENCE316} but is expected to deliver a higher power due to its continuous mode of operation (no adiabatic sweeps needed). We have presented possible implementations which employ Cooper pair boxes and flux qubits as working substances, two-qubit quantum gates, and on-chip fast calorimetry for the detection of single exchanged energy quanta. The proposed experiment would allow for the first fully stochastic characterisation of a quantum heat engine.

\section*{Acknowledgements}
This research was supported by the 7th European Community Framework Programme under grant agreements n. 623085 (MC-IEF-NeQuFlux) (M.C.), n. 600645 (IP-SIQS) (R.F.), n. 618074 (STREP-TERMIQ) (R.F.), n. 308850 (INFERNOS) (J.P.); by the Italian Ministry of Education University and Research under grant agreement n. MIUR-PRIN-2010LLKJBX (R.F.); by the Academy of Finland (projects 250280 and 272218) (J.P.); and by the COST action MP1209 ``Thermodynamics in the quantum regime''.

\section*{References}

\providecommand{\newblock}{}

\end{document}